\documentclass[a4paper,fleqn,usenatbib]{mnras}

\usepackage[T1]{fontenc}
\usepackage{ae,aecompl}
\usepackage{subfig}
\usepackage[usenames,dvipsnames,svgnames,table]{xcolor}

\usepackage{graphicx}	
\usepackage{amsmath}	
\usepackage{amssymb}	
\usepackage{longtable}

\title[\mbox{The Ly$\alpha$} fraction with MUSE]{The MUSE-Wide survey: A measurement of the \mbox{Ly$\alpha$} emitting fraction among $z>3$ galaxies}

\author[J. Caruana et al.]{
Joseph Caruana$^{1,2,3}$\thanks{E-mail: joseph.caruana@um.edu.mt},
Lutz Wisotzki$^{3}$,
Edmund Christian Herenz$^{4}$,
Josephine Kerutt$^{3}$, 
\newauthor Tanya Urrutia$^{3}$, Kasper Borello Schmidt$^{3}$, Rychard Bouwens$^{5}$, Jarle Brinchmann$^{5,6}$,
\newauthor Sebastiano Cantalupo$^{7}$, Marcella Carollo$^{7}$, Catrina Diener$^{8}$, Alyssa Drake$^{9}$, 
\newauthor Thibault Garel$^{9}$, Raffaella Anna Marino$^{7}$, Johan Richard$^{9}$, Rikke Saust$^3$,
\newauthor Joop Schaye$^{5}$, Anne Verhamme$^9$
\\
$^{1}$Department of Physics, University of Malta, Msida MSD 2080, Malta\\
$^{2}$Institute for Space Sciences \& Astronomy, University of Malta, Msida MSD 2080, Malta\\
$^{3}$Leibniz Institut f\"{u}r Astrophysik, An der Sternwarte 16, 14482 Potsdam, Germany\\
$^{4}$Department of Astronomy, Stockholm University, AlbaNova University Centre, SE-106 91, Stockholm, Sweden\\
$^{5}$Leiden Observatory, Leiden University, P.O. Box 9513, 2300 RA, Leiden, The Netherlands\\
$^{6}$Instituto de Astrof{\'\i}sica e Ci{\^e}ncias do Espaço, Universidade do Porto, CAUP, Rua das Estrelas, PT4150-762 Porto, Portugal\\
$^{7}$ETH Zurich, Institute for Astronomy, HIT J31.5, Wolfgang-Pauli-Strasse 27, 8093 Zurich, Switzerland\\
$^{8}$Institute of Astronomy, Madingley Road Cambridge, CB3 0HA, UK\\
$^{9}$Univ Lyon, Univ Lyon1, Ens de Lyon, CNRS, Centre de Recherche Astrophysique de Lyon UMR5574, F-69230, Saint-Genis-Laval, France
}

\date{Accepted XXX. Received YYY; in original form ZZZ}

\pubyear{2015}

\begin{document}
\label{firstpage}
\pagerange{\pageref{firstpage}--\pageref{lastpage}}
\maketitle

\begin{abstract}
We present a measurement of the fraction of Lyman $\alpha$ (Ly$\alpha$) emitters ($X_{\rm{Ly} \alpha}$) amongst \emph{HST} continuum-selected galaxies at $3<z<6$ with the Multi-Unit Spectroscopic Explorer (MUSE) on the VLT.  Making use of the first 24 MUSE-Wide pointings in GOODS-South, each having an integration time of 1 hour, we detect 100 Ly$\alpha$ emitters and find $X_{\rm{Ly} \alpha}\gtrsim0.5$ for most of the redshift range covered, with 29 per cent of the Ly$\alpha$ sample exhibiting rest equivalent widths (rest-EWs) $\leq$ 15\AA.  Adopting a range of rest-EW cuts (0 - 75\AA), we find no evidence of a dependence of $X_{\rm{Ly} \alpha}$ on either redshift or UV luminosity.
\end{abstract}




\section{Introduction}

Lyman $\alpha$ (Ly$\alpha$) emitters have been the subject of a large number of studies over the past several years.  The Ly$\alpha$ line often being the strongest emission line in the UV for star-forming galaxies (see, e.g., Peebles \& Partridge 1967, Amor\'{\i}n et al. 2017), it holds the answer to several pieces of information, most crucially the determination of the galaxies' redshift.  Key questions about these objects revolve around their masses, ages and dust extinction - and their relationship to continuum selected galaxies, particularly the extent to which such galaxies exhibit this emission line.  The fraction of Ly$\alpha$ emitters amongst continuum-detected sources, the Ly$\alpha$ emitter fraction, $X_{\rm{Ly} \alpha}$, is related to the underlying distribution of Ly$\alpha$ equivalent widths (EWs) amongst these objects, thus yielding additional information to that provided by Ly$\alpha$ emitter luminosity functions.  In recent years, the evolution of the Ly$\alpha$ fraction with redshift has also seen widespread use in probing the neutral HI fraction of the intergalactic medium at $z>6$ (e.g.\ Pentericci et al.\ 2011, Caruana et al.\ 2012, Schenker et al.\ 2012, Ono et al.\ 2012, Caruana et al.\ 2014) making accurate measurements of $X_{\rm{Ly} \alpha}$ at lower redshifts crucial.  The measurement of $X_{\rm{Ly} \alpha}$ also provides a point of reference for theoretical approaches that use it as an observational assessment of galaxy evolution and reionization models (e.g. Dayal et al. 2011, Forero-Romero et al.\ 2012, Garel et al. 2015 \& 2016, Kakiichi et al. 2016; for a review, see Dijkstra, 2014) with some models failing to reproduce aspects of the Ly$\alpha$ emitter population such as high EW (>100\AA) emitters. 

A better understanding of the nature of Ly$\alpha$ emitters requires homogeneous, statistically-significant surveys of these objects.  Several studies have been conducted in this vein, mostly employing multi-object spectrosopy (e.g.\ Stark et al. 2010, Mallery et al. 2012, Cassata et al. 2015) and narrow-band imaging (e.g. Rhoads et al.\ 2000, Ouchi et al.\ 2008).  Both of these approaches carry their own respective drawbacks.  Multi-object slit spectroscopy often entails choosing an observing setup that is configured (e.g. via the choice of an appropriate grism) to target the required wavelength space.  This, in turn, requires a pre-selected catalogue of sources whose probable redshift range has already been estimated (e.g. via photometric selection).  Moreover, the slit geometry can pose problems and necessitate compromises (e.g. with overlapping slits in the case of objects in spatial proximity) and the spectra themselves are prone to slit losses.  Finally, skyline contamination can be a greater problem compared to narrow-band searches.  On the other hand, in the case of narrowband imaging the setup is tuned to a specific redshift, meaning that this approach is more suited to surveying a narrow redshift slice.  Furthermore, the sensitivity can be lower than in the case of slit spectroscopy (by virtue of the filter width being wider than the spectral extent of the Ly$\alpha$ emission line).

The ESO-VLT Multi Unit Spectroscopic Explorer (MUSE, Bacon et al.\ 2010) is an integral field spectrograph that offers both high spatial resolution ($0.2\times0.2$ arcsec) and a wide spectral range from 4750\AA\, to 9300\AA.  This wide wavelength coverage translates into a possibility to investigate Ly$\alpha$ over a wide redshift range spanning $z=2.91 - 6.64$, which was one of the principal scientific drivers for the construction of the instrument.  Being an IFU imager/spectrograph, it does away with the requirement to set up multiple slits with associated flux losses.  These advantages together with its high throughput and wide field of view (1 arcmin$^{2}$) make MUSE an optimal instrument for Ly$\alpha$ surveys.  

In this study, we present results from MUSE-Wide, a relatively shallow survey with MUSE, taken as part of Guaranteed Time Observations (GTO).  Basing on \emph{HST} imaging catalogues (Guo et al. 2013, Skelton et al. 2014) for GOODS South, our study focuses on objects that are continuum-bright ($m_{775W} < 26.5$), and survey for relatively bright emission line galaxies with the emphasis lying on wide-area coverage whilst employing relatively short integration times.  This approach enables us to straightforwardly determine the redshift for all sources that exhibit Ly$\alpha$.  We investigate $X_{\rm{Ly} \alpha}$ and its relation to UV luminosity and redshift, and derive EW measurements for all emitters in our sample, which include a number of objects exhibiting very low (sub-10\AA) EWs.  Our results demonstrate the benefit of employing IFU spectroscopy combined with optimal spectral extraction (detailed in \ref{sec:spec_extraction}), which allows us to better capture the flux from sources that would otherwise go undetected.  This suggests that current estimates of the Ly$\alpha$ fraction might be underestimating the number of Ly$\alpha$ emitters, and by extension, inferences on the evolution of this fraction both with redshift and $M_{\rm{UV}}$ should be considered with caution.

In this paper, we adopt a $\Lambda$CDM cosmology throughout, with $\Omega_{M}=0.3$, $\Omega_{\Lambda}=0.7$ and $H_{0}=70$ kms$^{-1}$Mpc$^{-1}$.  Magnitudes are given in the AB system (Oke \& Gunn 1983).

\section{Observations and Data Processing}

The MUSE-Wide project (see also Herenz et al. 2017) is a blind spectroscopic survey (PI L. Wisotzki) using the MUSE panoramic integral field spectrograph at the ESO-VLT, carried out as a part of the GTO awarded to the MUSE consortium. The final survey covers some 100 arcmin$^2$ in areas with deep \emph{HST} imaging and complementary multi-wavelength data, with the Chandra Deep Field South (CDFS) as the primary region of interest. MUSE covers a fixed spectral range from 4750\AA\ to 9300\AA\ with a resolution of 2.5\AA\ (FWHM).

This paper uses data from the first 24 MUSE-Wide pointings in the CDFS-Deep part of the CANDELS \emph{HST} imaging survey (Cosmic Assembly Near-Infrared Deep Extragalactic Legacy Survey; Grogin et al. 2011, Koekemoer et al. 2011), which in turn was built on top of the earlier GOODS imaging campaign (Giavalisco et al. 2004). Our 24 MUSE fields are also all located within the footprint of the 3D-HST grism survey (Brammer et al. 2012, Momcheva et al. 2016). 

A detailed account of the observations, calibration and data reduction procedures will be given in a forthcoming publication (Urrutia et al., in preparation) accompanying the first data release of MUSE-Wide. In brief, all calibration exposures followed the ESO calibration plan. For the data reduction we used the MUSE data reduction pipeline (v1.0) with custom enhancements of the flat fielding and sky subtraction steps. After processing each single exposure separately and converting it into a datacube on a pre-defined world coordinate system grid, the four 900s exposures of one pointing were coadded into a final single datacube. These 24 datacubes, one for each MUSE pointing in the CDFS, were the basis of all further analysis.

\section{Analysis}
\subsection{Catalogues of \emph{HST} continuum-detected sources}
The CANDELS team produced a catalogue of all detected continuum sources in GOODS South (Guo et al.\ 2013).  We made use of this catalogue for the purpose of our study, applying a magnitude-cut to select all sources from this catalogue that satisfied $m_{775W}<26.5$.  Since the CANDELS/3D-HST NIR-detected catalogue by Skelton et al.\ (2014) also includes photometric redshifts, we cross-matched the Guo et al. (2013) catalogue to that of Skelton et al. (2014); this allowed us to also apply a photometric-redshift cut to our selection.  We adopted a conservative cut and selected those sources which satisfied $z_{\rm{phot}}>2$.  Such a low $z_{\rm{phot}}$ threshold (Ly$\alpha$ only enters the MUSE spectral range at $z>2.9$) was adopted in order to minimise the number of sources which potentially had a larger error on their photometric redshift.  At the same time, this choice sped up our analysis considerably by greatly cutting down on the number of sources whose spectra had to be subsequently inspected for \mbox{Ly$\alpha$}. These selection criteria resulted in a list of 579 sources, revised to 532 following the removal of 47 objects that lay close to the edge of the MUSE field-of-view.  To define ``edge-objects'', for each source we considered the cube layer at which \mbox{Ly$\alpha$} peaked (or, if no emission line was visible, would be expected to peak basing upon $z_{\rm{phot}}$) and checked whether there was any pixel within a spatial radius of 13 pixels (in that layer) that had less than 2 exposure cubes contributing to its value.  Such cases were defined to be edge-objects and removed from the list.

\subsection{Spectral extraction}
\label{sec:spec_extraction}
The light profile of each object as imaged with the \emph{HST} F775W filter was modelled with a 2D-Gaussian.  (That is, effectively, the \emph{HST} imaging data was used to provide us with a prior on the shape of the object.)  For any given target field, the MUSE Point Spread function (PSF) was fit using bright stars where these were available.  Where not, a number of these 2D-Gaussian galaxy models (around ten per field) were convolved with a varying set of Gaussians, and the results were subtracted from the MUSE images of the same galaxies. The best fitting (smallest residual) Gaussian function was taken to be the PSF.  The procedure was carried out over several spectral bins of the MUSE datacube to derive the wavelength dependence of the PSF.  (Further details on the PSF estimation are found in Herenz et al. 2017.)

\begin{figure}
\begin{center}
   \resizebox{0.5\textwidth}{!}{\includegraphics{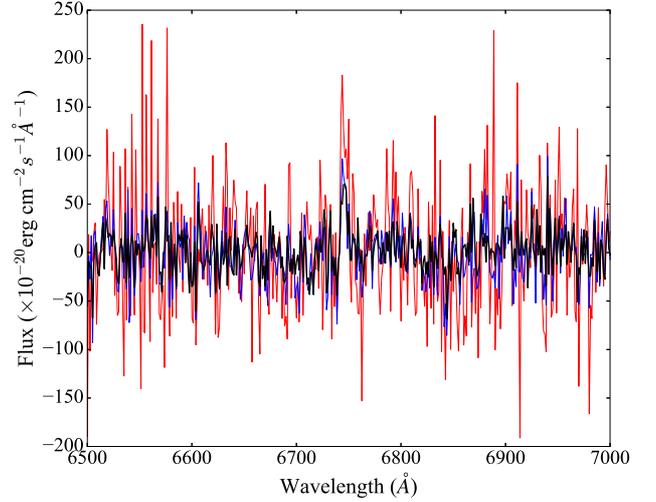}} \\
\end{center}
 \caption{The spectrum shown in black has been optimally extracted as described in Section \ref{sec:spec_extraction}, whereas circular aperture extractions with radii of 10 and 5 pixels are shown in red and blue respectively.  The optimal extraction improves the S/N ratio as it down-weights noisier pixels in the outer regions of the aperture.  The suppression of noisy spikes in the spectra greatly facilitates visual searches for Ly$\alpha$.}
 \label{fig:ap_vs_opt}
\end{figure}

Following this PSF estimation for each field, each 2D-Gaussian galaxy model was analytically convolved with the wavelength-dependent 2D-Gaussian model of the PSF.  The resulting MUSE PSF-convolved template was then used to optimally extract the spectra, where (following directly from the least-squares condition applied to the template matching problem), the flux in spectral layer $k$ is given by:

\begin{equation}
\label{eqn:horne}
\alpha_{k}=\frac{\sum_{ij}\left[(d_{ijk}\times t_{ijk})/\sigma_{ijk}^{2}\right]}{\sum_{ij}\left[t_{ijk}^{2}/\sigma_{ijk}^{2}\right]}
\end{equation}

where $d_{ijk}$, $t_{ijk}$ and $\sigma_{ijk}$ denote respectively the value of the data, template and standard deviation ($\sqrt{var}$) at the voxel with coordinates $i$,$j$,$k$.  An example spectrum extracted via this method is shown in Fig. \ref{fig:ap_vs_opt}.  The spectra that were extracted via this method were then searched for Ly$\alpha$ emission but were not used for flux measurements since while they exhibit an improved signal-to-noise ratio (S/N), the total flux can be biased.

\subsection{Assembling the \mbox{Ly$\alpha$} catalogue}
\label{sec:assembling_catalog}

To assemble our catalogue of \mbox{Ly$\alpha$} emitters we used a data product from the software \textsc{lsdcat}\footnote{The source code is available in Herenz \& Wisotzki (2016).} (Herenz \& Wisotzki 2017) to facilitate the search for emission lines in our spectra.  \textsc{lsdcat} is a tool that was developed to find line emitters which lack a continuum detection in MUSE datacubes.  Whilst we did not make use of \textsc{lsdcat}'s line-emitter cataloguing function, as we are here interested in sources which \emph{do} exhibit continuum emission in the \emph{HST} images (and, for the brighter objects, the MUSE cubes), one of its data products was useful for our analysis, as we describe next.  

The premise of \textsc{lsdcat} is based on matched-filtering, whereby the datacubes are cross-correlated with a template that represents an expected emission line's 3D profile, thus maximising the signal-to-noise of faint emission lines.  One of its data products is an S/N cube, every voxel of which represents the S/N of the respective voxel in a MUSE datacube.  An S/N spectrum from this S/N cube was extracted at the centre-coordinates of each source in our continuum-bright-selected catalogue.   We inspected this S/N spectrum to search for features with S/N $>4$, and the wavelength at which such peaks occurred was recorded.  Following this, for each recorded spectral feature we ran a search for a higher S/N peak in surrounding voxels in the cube (namely, within a circular radius of 3 voxels spatially and 1 voxel spectrally).  The reason for this procedure is to: (1) account for any discrepancies in astrometry, and (2) take into account the possibility that peak \mbox{Ly$\alpha$} emission may occur in a voxel that is spatially offset from the centre coordinates based on the continuum image (at which coordinates the S/N spectrum had been extracted), and which therefore might also exhibit a corresponding slight shift in the spectral direction.  Where this routine returned a higher S/N peak in a surrounding voxel, the spatial and spectral coordinates of this peak were recorded for that particular feature.  

The spectral resolution of MUSE is sufficient to resolve the separate components of the [O II] doublet, which greatly reduces the possibility of mistaking [O II] for Ly$\alpha$.  We visually inspected the spectral features in the (non-smoothened) optimally-extracted spectra.  We also inspected the shape of the spectral features following a simple, circular, varying-size aperture extraction to further guard against mistaking artefacts for genuine emitters.  As a final check, for each spectral feature we also inspected the layer in the cube where the peak of the spectral feature occurred.  A cosmic ray hit or other artefact would in general be expected to exhibit very narrow spatial extent and is easier to flag in a 2D cube layer.  In the end, following this visual inspection, for our final \mbox{Ly$\alpha$} catalogue we utilised an S/N cut of 5.0, as this was determined to securely guard against artefacts.  Following further work on assessing the MUSE datacubes' noise properties, it was found that the effective noise was initially underestimated by a factor of 1.2, which effectively means that our S/N=5.0 cut actually corresponds to S/N$\approx4.0$ and we are able to distinguish real emitters from artefacts at this lower S/N level.  Ly$\alpha$ emission was securely detected in 100 sources in our sample.

\subsection{Redshift determination}

We define the Ly$\alpha$ emitter fraction, $X_{\rm{Ly} \alpha}$ as:

\begin{equation}
\label{eqn:fraction}
X_{\rm{Ly} \alpha} = \frac{\mbox{\emph{HST} continuum-detected sources exhibiting Ly } \alpha}{\mbox{\emph{HST} continuum-detected sources}}
\end{equation}

Prior to making any further use of the photometric redshifts - essential for the determination of the denominator in Equation \ref{eqn:fraction} - we applied a redshift correction, the motivation for which is described below.

\begin{figure}
\begin{center}
   \resizebox{0.5\textwidth}{!}{\includegraphics{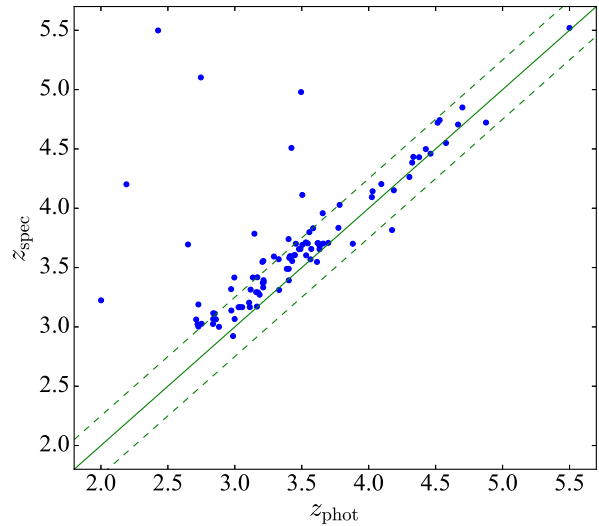}} \\
\end{center}
 \caption{Spectroscopic redshift vs. photometric redshift.  The two dashed lines correspond to $|z_{\rm{spec}}-z_{\rm{phot}}|=0.25$.  The average difference between $z_{\rm{spec}}$ and $z_{\rm{phot}}$ of $0.1$ ($\pm 0.01$) was used as a correction factor for the photometric redshifts.}
 \label{fig:specz_vs_photz}
\end{figure}

Following the identification of Ly$\alpha$ emitters in our sample, we investigated the relation between photometric redshift and spectroscopic redshift for the entire Ly$\alpha$ sample, as shown in Fig. \ref{fig:specz_vs_photz}.  As is evident from this figure, with the exception of 10 objects, all sources exhibit a spectroscopic redshift that is slightly higher than the corresponding photometric redshift found by Skelton et al. (2014), an effect that is also visible in figure 23 of the same paper for this redshift range.  Oyarz\'{u}n et al. (2016) found that the magnitude of this offset correlates with Ly$\alpha$ EW, so a possible source for this discrepancy could be due to the Ly$\alpha$ emission line altering the photometry, an effect that is not accounted for in the photo-z SED templates (see Schaerer \& de Barros 2012).  A detailed investigation of this systematic offset between photometric and spectroscopic redshifts will be presented in Brinchmann et al. (submitted), where the role of the applied intergalactic absorption model and the effect of spatially overlapping galaxies are also explored.

\begin{figure}
\begin{center}
   \resizebox{0.5\textwidth}{!}{\includegraphics{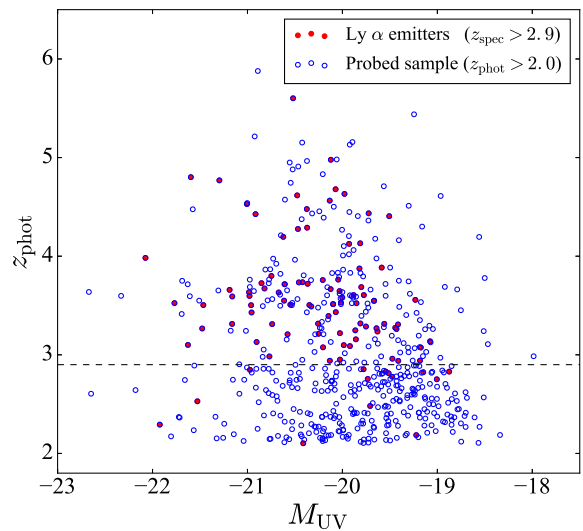}} \\
\end{center}
 \caption{The sample of 100 Lyman $\alpha$ emitting galaxies (red) amongst the entire sample of 532 continuum-selected objects with $z_{\rm{phot}}>2.0$ (open blue circles).  The dashed horizontal line marks $z=2.9$, the redshift at which Ly$\alpha$ enters the MUSE redshift window.}
 \label{fig:sample}
\end{figure}

Ignoring outliers on this plot, where by outliers we mean sources for which $|z_{\rm{spec}}-z_{\rm{phot}}|>0.25$ (represented in Fig. \ref{fig:specz_vs_photz} by two dashed lines), we calculate the required redshift correction, calculated as the mean of the difference between $z_{\rm{phot}}$ and $z_{\rm{spec}}$ for each Ly$\alpha$ emitter:

\begin{equation}
\label{eqn:error}
\mbox{Redshift correction}=\frac{1}{N}\sum_{i}^{N}({z_{\rm{phot}_{i}}-z_{\rm{spec}_{i}})}
\end{equation}

where $N$ denotes the total number of Ly$\alpha$ emitters and $i$ is an individual emitter.  We find this redshift correction to be 0.10 ($\pm0.01$).  We correct the photometric redshift of the entire sample of 532 objects.  Out of 198 continuum sources with $z_{\rm{phot}}$ (or $z_{\rm{spec}}$ where available) > 2.9 (which is the redshift at which Ly$\alpha$ enters the MUSE redshift range) and -22.5 < $M_{\rm{UV}}$ < -18.5, we find Ly$\alpha$ emission in 100 sources (see Fig. \ref{fig:sample}).

\subsection{Flux measurements}

The \textsc{lsdcat} routine \texttt{lsd\_cat\_measure.py} was used to measure the fluxes of the Ly$\alpha$ lines. This routine creates a `pseudo-narrow' band image from the datacube centered on the emission line.  The bandwidth of this image is defined by the spectral layers in which the emission line is above a certain analysis threshold $\mathrm{S/N}_\mathrm{ana.}$ in the S/N cube.  By visual inspection (separate from that described in Sec. \ref{sec:assembling_catalog}), we found that $\mathrm{S/N}_\mathrm{ana.} = 4$ separates the emission line signal from the noise.  The flux is then integrated in these narrow-band images within $3\times R_\mathrm{Kron}$ apertures, where $R_\mathrm{Kron}$ is the characteristic light distribution weighted radius introduced by Kron (1980), centered on the first central-moment calculated in a PSF-smoothed version of the pseudo-narrow image.  In the vast majority of cases, \textsc{lsdcat}'s automatic line flux measurements agree well with fluxes determined from a manual curve-of-growth analysis (Herenz \& Wisotzki 2017, accepted).

\subsection{EW measurements}

For most sources, the 1 hour integration time with MUSE was not sufficient to detect the continuum directly from the extracted MUSE spectra, so we used \emph{HST} ACS imaging in the F814W band to obtain the continuum flux density for our sources.  The F814W band is the deepest \emph{HST} band for CANDELS CDFS and therefore best suited for this purpose.  Whilst Ly$\alpha$ enters the redshift window of the F814W band at z=4.97-6.23, we note that only four objects in the entire Ly$\alpha$ sample lie in this redshift range.  Moreover, even in such cases, Ly$\alpha$ contamination is not expected to have any significant effect on the F814W magnitude since: (a) the galaxies are selected to be continuum-bright, and (b) the scale-length of Ly$\alpha$ is much larger than the size of the aperture used for \emph{HST} flux measurements, meaning that much of the Ly$\alpha$ emission is excluded. Therefore, any subtraction of the Ly$\alpha$ flux from the F814W magnitude would almost certainly result in over-subtraction.  

We used \texttt{GALFIT} (Peng et al. 2010) to fit each of our objects with Sersic profiles to optimally determine their magnitudes.  In converting the magnitude to continuum flux density at the position of the Ly$\alpha$ line, we assumed a continuum slope with a mean $\beta=-2$ (e.g. Ouchi et al. 2008, Blanc et al. 2011, Castellano et al. 2014).  From this measure of the continuum and the line flux measurements obtained directly from the MUSE datacubes (via \textsc{lsdcat}), we computed the EWs for our sources.  

Traditionally, both the line flux and the continuum flux are measured in the same fixed aperture.  This, however, is not optimal because there is evidence that Ly$\alpha$ emission is more extended than the UV continuum (Xue et al. 2017, Wisotzki et al. 2016, Momose et al. 2016). This would necessitate the use of larger apertures for the line flux measurements.  However, two problems arise if one were to use the same larger aperture to also obtain a measurement of the continuum flux density.  Firstly, the noise would increase and, secondly, the large size of the aperture could potentially include other sources (e.g. low-redshift interlopers).  Therefore, taking advantage of the deep broadband data and deriving individual fits for objects is the best approach for the present sample of Ly$\alpha$ emitting galaxies.

\section{Results \& Discussion}

The final set of Ly$\alpha$ emitting galaxies consists of 100 objects spanning $z=2.92 - 5.52$ (see Table \ref{table:sample}).  

\begin{figure}
\begin{center}
   \resizebox{0.5\textwidth}{!}{\includegraphics{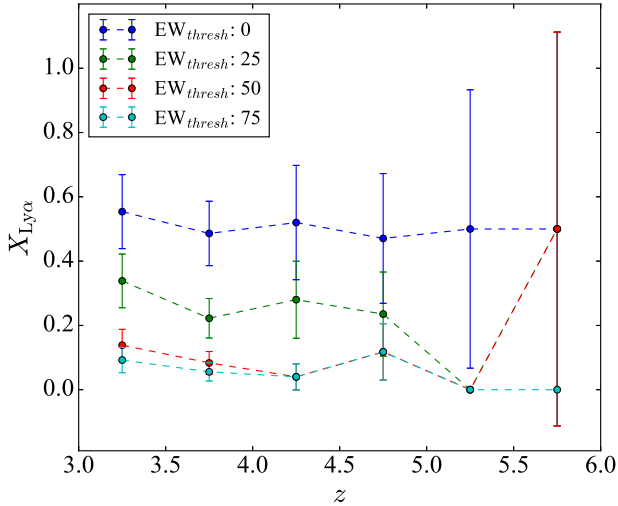}} \\
\end{center}
 \caption{$X_{\rm{Ly} \alpha}$ vs. redshift for different rest-EW thresholds.  The error bars are derived via error-propagation of Poissonian statistics, such that $\sigma=(N_{\rm{Ly} \alpha}/N_{z}^{2}+N_{\rm{Ly} \alpha}^{2}/N_{z}^{3})^{1/2}$ where $N_{\rm{Ly} \alpha}$ is the number of Ly$\alpha$ emitters within a given redshift bin and $N_{z}$ is the number of continuum sources in the same bin.}
 \label{fig:frac_vs_z}
\end{figure}

Considering Figure \ref{fig:frac_vs_z}, which focuses on $3.0<z<6.0$\footnote{By virtue of the redshift range selected (which allows for straightforwardly-cut redshift bins), this figure omits 5 emitters with $2.9<z<3.0$.}, we find $X_{\rm{Ly} \alpha}\approx0.5$ or larger over this redshift range when the entire sample of Ly$\alpha$ emitters is considered (see also Table \ref{table:xlya_vs_z}).  We further consider $X_{\rm{Ly} \alpha}$ for three rest-EW cuts: 25\AA, 50\AA, and 75\AA.  Comparing with previous studies, we note that Cassata et al.\ (2015) find $X_{\rm{Ly} \alpha}\approx0.12$ over $z=3-4$ for rest-EW $> 25$\AA.  For the same EW cut, we find $X_{\rm{Ly} \alpha}\geq0.22$, a two-fold increase in the fraction of Ly$\alpha$ emitters in this same redshift range.  

We investigated the variation of $X_{\rm{Ly} \alpha}$ with $z$, adopting Poissonian statistics for our error bars such that the propagated error for $X_{\rm{Ly} \alpha}$ is $\sigma=(N_{\rm{Ly} \alpha}/N_{z}^{2}+N_{\rm{Ly} \alpha}^{2}/N_{z}^{3})^{1/2}$.  Formalising our null hypothesis to state that there is no correlation between $X_{\rm{Ly} \alpha}$ and $z$, we perform weighted least squares regression on the data for each rest-EW cut, and derive F-test $p$-values of 0.13 (0\AA), 0.39 (25\AA), 0.22 (50\AA) and 0.35 (75\AA), all falling short of the 95 per cent confidence level ($p=0.05$).  Therefore, even adopting this simplified (propagation of $\sqrt N$ error) approach, we fail to reject the null hypothesis, discerning no dependence of $X_{\rm{Ly} \alpha}$ on $z$ regardless of the EW cut adopted. While data points at $z>5$ have relatively large uncertainties, potentially obscuring an underlying trend of increasing $X_{\rm{Ly} \alpha}$, no dependence of $X_{\rm{Ly} \alpha}$ on $z$ is observed at $z<5.0$ either.  At any rate, this exercise suggests that any underlying trend cannot be particularly strong.

\begin{figure*}
    \centering
    \subfloat[$X_{\rm{Ly} \alpha}$ vs. UV luminosity for various rest-EW cuts. We find no evidence of a correlation between the two quantities for any of the rest-EW cuts adopted.  The error bars are derived via error-propagation of Poissonian statistics, such that $\sigma=(N_{\rm{Ly} \alpha}/N_{M_{\rm{UV}}}^{2}+N_{\rm{Ly} \alpha}^{2}/N_{M_{\rm{UV}}}^{3})^{1/2}$ where $N_{\rm{Ly} \alpha}$ is the number of Ly$\alpha$ emitters within a given $M_{\rm{UV}}$ bin and $N_{M_{\rm{UV}}}$ is the total number of continuum sources in the same bin.]{{\includegraphics[width=8cm]{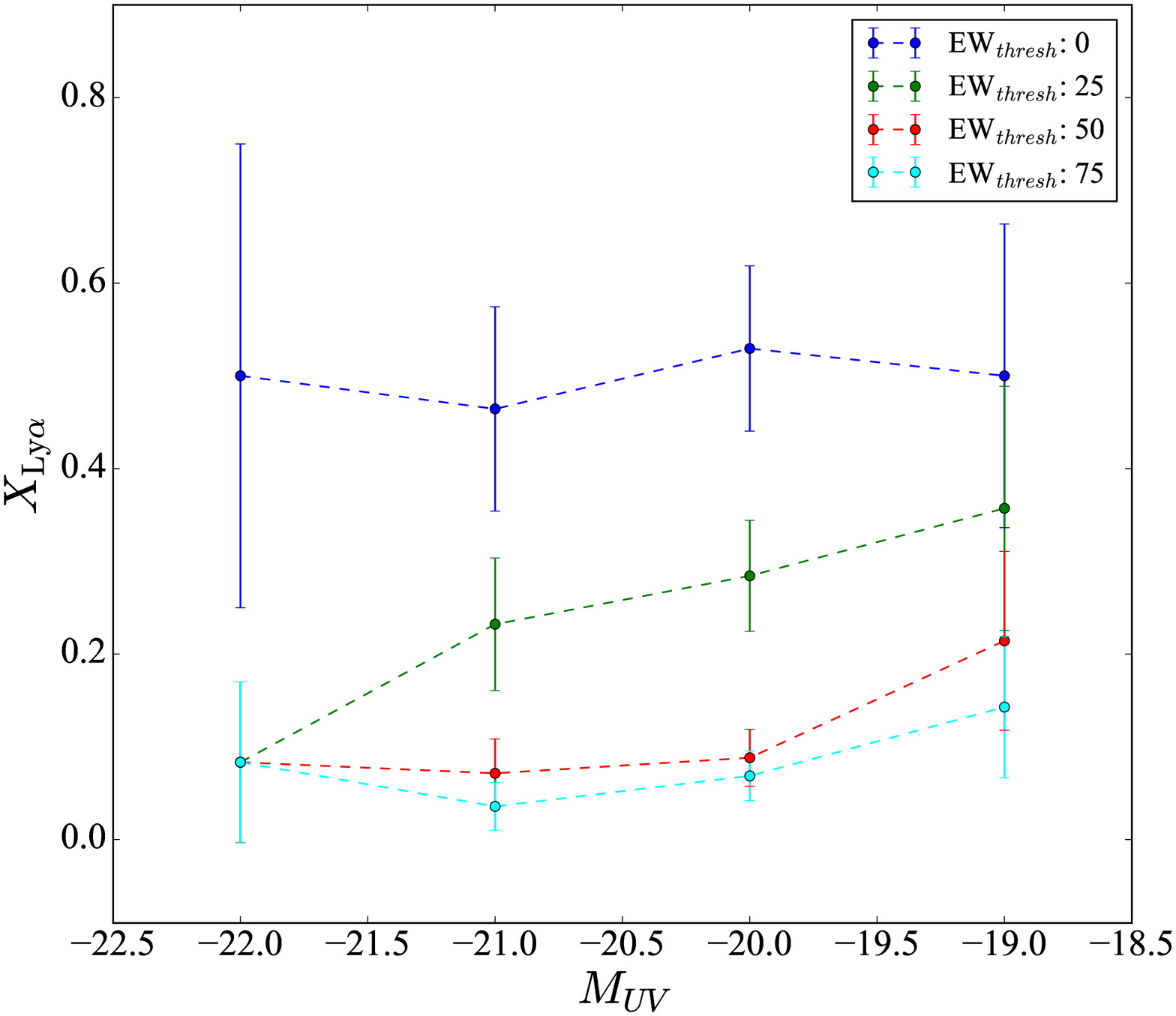} }}%
    \qquad
    \subfloat[rest-EW vs. $M_{\rm{UV}}$. There is a smaller number of bright (M$_{UV}\lesssim-21.0$) objects with moderate to high Ly$\alpha$ EWs. (Note that this figure omits 3 outlying sources with very large rest-EWs (366\AA,\ 633\AA\ and 1118\AA\ at $M_{\rm{UV}}= $ -20.48, -21.77 and -19.01 respectively) in order to aid better viewing of the rest of the sample.)]{{\includegraphics[width=8cm]{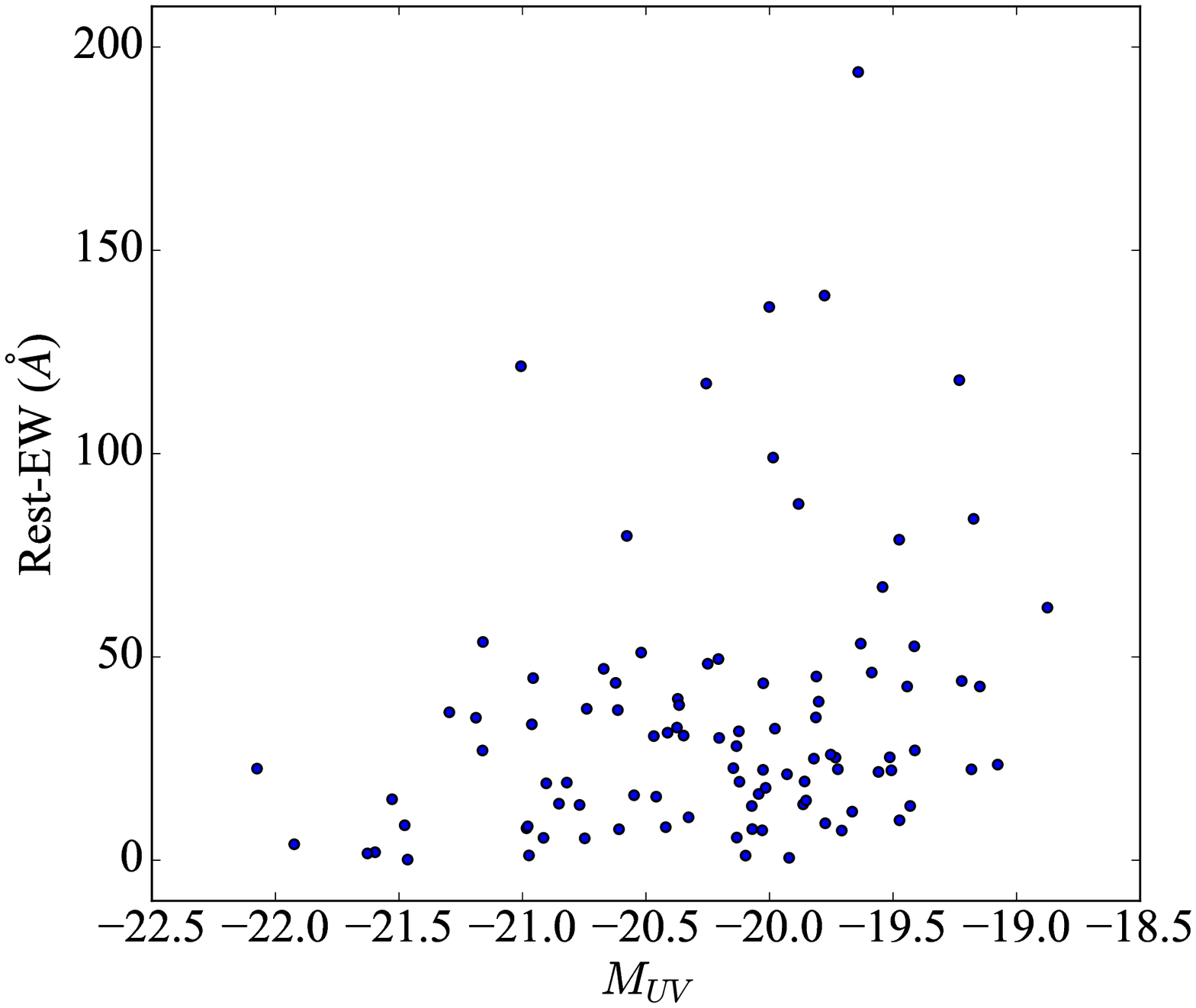} }}%
    \caption{$X_{\rm{Ly} \alpha}$ vs. UV luminosity and rest-EW vs. $M_{\rm{UV}}$}%
    \label{fig:frac_vs_absmag}%
\end{figure*}

We also investigated any potential trends of $X_{\rm{Ly} \alpha}$ with UV luminosity, again adopting the same rest-EW cuts and employing the above analysis, finding $p=0.46$ (0\AA), 0.04 (25\AA), 0.37 (50\AA), and 0.27 (75\AA).  Some previous studies note a trend of a rising $X_{\rm{Ly} \alpha}$ with fainter UV luminosity (e.g.\ Stark et al.\ 2010 for $z=3.5-6$).  However, as shown in Fig. \ref{fig:frac_vs_absmag}a, our first MUSE-Wide results do not seem to indicate any overall significant correlation between the two quantities except marginally for the case with a rest-EW cut of 25\AA\, (p=0.04), thus being more in line with the findings of Cassata et al. (2015), who also do not find such a relationship.  It will be interesting to explore in subsequent studies whether a larger sample will reveal any dependence of $X_{\rm{Ly} \alpha}$ on $M_{\rm{UV}}$.

29 per cent of the Ly$\alpha$ sample exhibits rest-EWs $\leq$ 15\AA.  We should note that our sample is S/N-limited, which effectively translates into it being flux-limited at a given redshift.  The S/N provided by \textsc{lsdcat} is dependent on the compact Ly$\alpha$ flux, which may be (albeit not necessarily) smaller than the total flux.  This is attributable to there being: (a) a range of halo sizes and line-widths, and (b) an error on the flux measurement itself.  Effectively, this means that a given S/N value corresponds to a range of measured fluxes.  For a continuum magnitude-defined sample, this flux limit (picked arbitrarily from the selection function at 50 per cent completeness; see Herenz et al., in preparation, for details) can be converted into an EW limit.  Across the redshift range explored by our data, we find median rest-EW limits of 30.36 ($z=3.0-3.5$), 27.86 ($z=3.5-4.0$), 25.63 ($z=4.0-4.5$), 23.22 ($z=4.5-5.0$), 35.19 ($z=5.0-5.5$) and 37.73 ($z=5.5-6.0$), with the overall median rest-EW in the $z=3.0 - 5.5$ range being 27.86\AA.  Such a limit, however, is strongly dependent on the continuum magnitude.  This, in fact, explains the existence of very low EW emitters in our sample - smaller even than what one would otherwise expect; even a small EW emitter will be detected if the source has a bright enough continuum.  The probing of $\lesssim 10$\AA\, EWs highlights the excellent capability of MUSE to detect low rest-EW sources with relatively short (i.e. 1 hour) integration times.  By adding sensitivity to the low-EW regime, even `shallow' surveys with MUSE can provide new insight into the size of the overall fraction of Ly$\alpha$ emitters.

This work also raises the possibility of implications for higher redshift observations.  Presently, various studies in the literature observe a drop in $X_{\rm{Ly} \alpha}$ at higher redshifts (e.g. \ Caruana et al.\ 2014, Tilvi et al. 2014, Treu et al. 2013).  Given the flat trend of $X_{\rm{Ly} \alpha}$ with redshift observed in this study, such a result could suggest that there might be a significant component of extended Ly$\alpha$ residing in the halos surrounding these sources (see also Wisotzki et al.\ 2016) which could have been missed by previous studies utilising slit spectroscopy by virtue of the slit not being large enough to encompass this emission. Furthermore, since Ly$\alpha$ emission with a high rest-EW seems to be more readily observable amongst fainter galaxies (Fig. \ref{fig:frac_vs_absmag}b), the search for this line might be more fruitful were one to look at fainter objects rather than targeting the brighter galaxies.  However, at present, this remains an open question; if the universe has a significant HI neutral fraction at $z>6$, then lower-luminosity galaxies might not be able to ionize a sufficiently large HII bubble around them, which would result in stronger attenuation of Ly$\alpha$ in these fainter systems.

This study illustrates that MUSE can probe low EW emitters at redshifts up to $z\approx5$ with relatively short integration times.  With integrations of the order of 1 hour, we are able to probe sub--10\AA\ EWs - and can detect Ly$\alpha$ emitters with rest-EW $>10$\AA,\ constituting $\approx80$ per cent of our sample, with high confidence.

\definecolor{verylightgray}{gray}{0.95}
\begin{table*}
\setlength\extrarowheight{3pt}
\centering
\begin{tabular}{| l | r | r | r | r | r | r |}
\hline
 & $3.0 < z < 3.5$ & $3.5 < z < 4.0$ & $4.0 < z < 4.5$ & $4.5 < z < 5.0$ & $5.0 < z < 5.5$ & $5.5 < z < 6.0$ \\ 
 \hline
 \hline
 \rowcolor{verylightgray}
EW$_{thresh}=0$\AA    & 36 / 65  &  35 / 72 &  13 / 25 &  8 / 17 & 2/4 & 1/2\\  
EW$_{thresh}=25$\AA  & 22 / 65  &  16 / 72 &  7 / 25 &  4 / 17 & 0/4 & 1/2\\
\rowcolor{verylightgray}
EW$_{thresh}=50$\AA  & 9 / 65  &  6 / 72   &  1 / 25   &  2 / 17  & 0/4 & 1/2 \\
EW$_{thresh}=75$\AA  & 6 / 65  &  4 / 72   &  1 / 25   &  2 / 17  & 0/4 & 0/2\\
\hline
 \end{tabular}
 \caption{The Ly$\alpha$ fraction per redshift bin.  In each case, the numerator denotes the number of sources exhibiting Ly$\alpha$ emission within a given redshift bin, whereas the denominator denotes the number of continuum sources in the same bin.  These fractions are represented in Figure \ref{fig:frac_vs_z}.}
  \label{table:xlya_vs_z}
\end{table*}

\definecolor{verylightgray}{gray}{0.95}
\begin{table*}
\setlength\extrarowheight{3pt}
\centering
\begin{tabular}{| l | r | r | r | r |}
\hline
& $-22.5 < M_{\rm{UV}} < -21.5$ & $-21.5 < M_{\rm{UV}} < -20.5$ & $-20.5 < M_{\rm{UV}} < -19.5$ & $-19.5 < M_{\rm{UV}} < -18.5$ \\ 
 \hline
 \hline
 \rowcolor{verylightgray}
EW$_{thresh}=0$\AA   & 6 / 12  &  26 / 56 &  54 / 102 &  14 / 28\\  
EW$_{thresh}=25$\AA & 1 / 12  &  13 / 56 &  29 / 102 &  10 / 28\\
\rowcolor{verylightgray}
EW$_{thresh}=50$\AA & 1 / 12  &  4 / 56   &  9 / 102 &  6 / 28\\
EW$_{thresh}=75$\AA & 1 / 12  &  2 / 56   &  7 / 102   &  4 / 28\\
\hline
 \end{tabular}
 \caption{The Ly$\alpha$ fraction per magnitude bin.  In each case, the numerator denotes the number of sources exhibiting Ly$\alpha$ emission within a given $M_{\rm{UV}}$ bin, whereas the denominator denotes the total number of continuum sources in that bin.  These fractions are represented in Figure \ref{fig:frac_vs_absmag}a.}
  \label{table:xlya_vs_mag}
\end{table*}

\section*{Acknowledgements}

This research is based on observations collected at the European Organisation for Astronomical Research in the Southern Hemisphere under ESO programme 094.A-0205(B).  

We thank the anonymous referee whose helpful comments greatly improved this manuscript.

We acknowledge funding by the Competitive Fund of the Leibniz Association through grants SAW-2013-AIP-4 and SAW-2015-AIP-2.  This work is supported by Fundação para a Ciência e a Tecnologia (FCT) through national funds (UID/FIS/04434/2013) and by FEDER through COMPETE2020 (POCI-01-0145-FEDER-007672). During part of this work, JB was supported by FCT through Investigador FCT contract IF/01654/2014/CP1215/CT0003.   RAM acknowledges support by the Swiss National Science Foundation.  JS acknowledges ERC Grant agreement 278594-GasAroundGalaxies.  TG is grateful to the LABEX Lyon Institute of Origins (ANR-10-LABX-0066) of the Université de Lyon for its financial support within the program ``Investissements d'Avenir'' (ANR-11-IDEX-0007) of the French government operated by the National Research Agency (ANR).







\appendix

\section{The Ly$\alpha$ catalog}

\begin{table*}
\footnotesize
\setlength\extrarowheight{3pt}
\centering

\begin{tabular}{| c  c  c  c  c  c  c  c |}
\hline

ID & ID & $M_{UV}$ & Flux [3-Kron] & Flux $\sigma$ & rest-EW & rest-EW $\sigma$ & z \\ 
(Skelton) & (Guo) & (F814W) &  ($\times 10^{-20}$ erg cm$^{-2}$ s$^{-1}$)  &  ($\times 10^{-20}$ erg cm$^{-2}$ s$^{-1}$)  & (\AA) & (\AA) &   \\ 
\rowcolor{verylightgray}
\hline
\hline
\rowcolor{verylightgray}
18198 & 8932 & -21.77 & 114164.98 & 338.13 & 632.61 & 15.4 & 4.51 \\ 
18702 & 9262 & -20.9 & 4988.66 & 452.38 & 18.91 & 1.87 & 3.17 \\ 
\rowcolor{verylightgray}
18439 & 9093 & -20.74 & 3262.53 & 341.16 & 37.24 & 3.91 & 3.37 \\ 
14173 & 6905 & -20.75 & 680.32 & 184.99 & 5.38 & 1.46 & 3.71 \\ 
\rowcolor{verylightgray}
7781 & 4229 & -20.58 & 9534.21 & 601.57 & 79.76 & 6.95 & 3.2 \\ 
23859 & 12329 & -21.16 & 10591.73 & 574.63 & 53.67 & 2.95 & 3.66 \\ 
\rowcolor{verylightgray}
22379 & 11427 & -20.26 & 12063.1 & 548.33 & 117.24 & 5.44 & 3.39 \\ 
18974 & 9435 & -20.82 & 3059.63 & 354.4 & 19.09 & 2.36 & 3.66 \\ 
\rowcolor{verylightgray}
21324 & 10812 & -20.67 & 6175.14 & 460.42 & 47.06 & 3.61 & 3.71 \\ 
25614 & 13375 & -21.6 & 249.07 & 113.65 & 1.96 & 0.89 & 4.85 \\ 
\rowcolor{verylightgray}
20768 & 10433 & -21.46 & 42.13 & 2256.53 & 0.15 & 7.91 & 3.49 \\ 
13283 & 6531 & -22.07 & 4565.68 & 375.97 & 22.52 & 1.94 & 3.7 \\ 
\rowcolor{verylightgray}
11864 & 5783 & -20.98 & 1497.97 & 243.61 & 7.86 & 1.3 & 3.6 \\ 
12589 & 6235 & -20.55 & 2100.37 & 192.16 & 15.98 & 1.55 & 3.58 \\ 
\rowcolor{verylightgray}
15002 & 7233 & -20.85 & 1637.24 & 289.11 & 13.9 & 2.91 & 3.7 \\ 
14982 & 7259 & -21.48 & 3425.0 & 423.31 & 8.59 & 1.07 & 3.17 \\ 
\rowcolor{verylightgray}
17777 & 8701 & -21.16 & 6895.07 & 458.29 & 26.98 & 2.11 & 3.33 \\ 
16007 & 7775 & -20.91 & 654.41 & 158.68 & 5.49 & 1.33 & 4.38 \\ 
\rowcolor{verylightgray}
15158 & 7350 & -20.35 & 3196.17 & 359.63 & 30.65 & 3.53 & 3.39 \\ 
16710 & 8108 & -20.77 & 3532.28 & 340.15 & 13.59 & 1.71 & 3.0 \\ 
\rowcolor{verylightgray}
18517 & 9113 & -20.42 & 884.34 & 201.51 & 8.12 & 1.85 & 3.68 \\ 
17484 & 8544 & -20.96 & 1316.61 & 247.48 & 33.42 & 6.31 & 3.74 \\ 
\rowcolor{verylightgray}
15419 & 7464 & -21.92 & 380.46 & 137.68 & 3.92 & 1.46 & 4.2 \\ 
16523 & 8005 & -21.19 & 6613.16 & 307.01 & 35.04 & 2.0 & 3.8 \\ 
\rowcolor{verylightgray}
16492 & 7986 & -21.3 & 2031.47 & 221.07 & 36.38 & 4.02 & 4.71 \\ 
17539 & 8584 & -20.61 & 5504.49 & 325.41 & 36.93 & 3.52 & 3.61 \\ 
\rowcolor{verylightgray}
18429 & 9060 & -20.96 & 6883.15 & 329.17 & 44.79 & 2.43 & 3.57 \\ 
18384 & 9109 & -21.63 & 1126.04 & 294.31 & 1.67 & 0.51 & 3.07 \\ 
\rowcolor{verylightgray}
18841 & 9317 & -20.61 & 988.01 & 1802.64 & 7.61 & 13.89 & 3.55 \\ 
19906 & 9945 & -21.01 & 8841.52 & 356.03 & 121.5 & 9.24 & 4.5 \\ 
\rowcolor{verylightgray}
21106 & 10675 & -20.37 & 1985.24 & 285.89 & 32.62 & 4.74 & 4.43 \\ 
20804 & 10491 & -20.04 & 1307.61 & 254.49 & 16.29 & 3.33 & 3.7 \\ 
\rowcolor{verylightgray}
21734 & 11040 & -20.07 & 642.29 & 141.26 & 13.33 & 3.0 & 4.55 \\ 
23169 & 11909 & -20.12 & 1029.91 & 179.07 & 19.3 & 3.97 & 4.72 \\ 
\rowcolor{verylightgray}
15660 & 7587 & -19.73 & 2118.74 & 365.79 & 25.25 & 4.37 & 2.98 \\ 
12277 & 6113 & -19.86 & 739.76 & 2162.86 & 13.73 & 40.16 & 3.79 \\ 
\rowcolor{verylightgray}
12145 & 6060 & -19.82 & 1036.77 & 227.71 & 24.98 & 5.55 & 3.83 \\ 
14405 & 6983 & -19.77 & 417.6 & 114.95 & 9.08 & 2.51 & 4.11 \\ 
\rowcolor{verylightgray}
13558 & 6622 & -19.59 & 1778.54 & 234.15 & 46.16 & 6.31 & 4.03 \\ 
13253 & 6490 & -19.18 & 1233.8 & 1810.08 & 22.35 & 32.79 & 3.11 \\ 
\hline
 \end{tabular}
 \caption{The catalog of Ly$\alpha$ emitters.}
  \label{table:sample}
\end{table*}

\begin{table*}
\contcaption{}
\setlength\extrarowheight{3pt}
\centering
\begin{tabular}{| c | c | c | c | c | c | c | c | c |}
\hline

ID & ID & $M_{UV}$ & Flux [3-Kron] & Flux $\sigma$ & rest-EW & rest-EW $\sigma$ & z \\ 
(Skelton) & (Guo) & (F814W) &  ($\times 10^{-20}$ erg cm$^{-2}$ s$^{-1}$)  &  ($\times 10^{-20}$ erg cm$^{-2}$ s$^{-1}$)  & (\AA) & (\AA) &   \\ 
\hline
\hline

\rowcolor{verylightgray}
10447 & 5354 & -19.8 & 2096.78 & 315.5 & 39.0 & 5.94 & 3.83 \\ 
10849 & 5504 & -19.22 & 1234.62 & 234.96 & 44.08 & 8.63 & 2.98 \\ 
\rowcolor{verylightgray}
22284 & 11369 & -20.41 & 3032.79 & 387.22 & 31.36 & 4.01 & 3.22 \\ 
23895 & 12341 & -20.0 & 10591.73 & 574.63 & 136.07 & 7.49 & 3.66 \\ 
\rowcolor{verylightgray}
23150 & 11872 & -20.21 & 2101.3 & 238.72 & 49.48 & 6.3 & 3.96 \\ 
23881 & 12313 & -20.48 & 10591.73 & 574.63 & 365.81 & 28.33 & 4.72 \\ 
\rowcolor{verylightgray}
20679 & 10410 & -20.02 & 1291.02 & 214.98 & 17.79 & 3.0 & 3.56 \\ 
18978 & 9384 & -19.56 & 1141.04 & 207.66 & 21.71 & 4.0 & 3.56 \\ 
\rowcolor{verylightgray}
15549 & 7493 & -20.98 & -3548.04 & 2679.83 & 8.32 & 3.95 & 4.98 \\ 
19097 & 9462 & -20.25 & 3904.14 & 453.15 & 48.32 & 6.21 & 3.17 \\ 
\rowcolor{verylightgray}
16981 & 8268 & -20.13 & 1208.13 & 201.57 & 28.06 & 4.69 & 4.46 \\ 
15294 & 7416 & -20.15 & 2454.19 & 264.42 & 22.64 & 2.62 & 3.59 \\ 
\rowcolor{verylightgray}
15815 & 7663 & -20.37 & 2812.89 & 322.6 & 39.68 & 5.28 & 4.15 \\ 
15601 & 7570 & -19.75 & 1901.85 & 305.7 & 25.97 & 4.18 & 3.27 \\ 
\rowcolor{verylightgray}
14421 & 7043 & -19.86 & 1133.53 & 215.89 & 19.36 & 3.71 & 3.17 \\ 
13851 & 6766 & -19.81 & 1945.14 & 291.95 & 45.18 & 6.83 & 3.38 \\ 
\rowcolor{verylightgray}
14403 & 7004 & -19.08 & 1145.76 & 214.62 & 23.51 & 4.99 & 3.17 \\ 
14204 & 6920 & -19.98 & 1229.0 & 257.03 & 32.35 & 6.97 & 4.74 \\ 
\rowcolor{verylightgray}
11528 & 5787 & -20.37 & 3104.33 & 380.56 & 38.16 & 6.32 & 3.71 \\ 
9990 & 5164 & -19.99 & 6221.49 & 574.65 & 99.03 & 9.85 & 3.42 \\ 
\rowcolor{verylightgray}
9553 & 5004 & -19.63 & 3703.81 & 401.12 & 53.27 & 5.85 & 3.42 \\ 
8885 & 4733 & -19.81 & 2507.22 & 325.77 & 35.11 & 32.74 & 4.14 \\ 
\rowcolor{verylightgray}
23111 & 11857 & -20.62 & 3949.69 & 411.04 & 43.61 & 4.72 & 4.2 \\ 
13755 & 6717 & -19.41 & 1641.82 & 332.96 & 27.01 & 5.57 & 3.02 \\ 
\rowcolor{verylightgray}
15282 & 7400 & -19.47 & 714.8 & 251.45 & 9.82 & 3.53 & 3.06 \\ 
15632 & 7566 & -19.67 & 746.34 & 240.56 & 11.94 & 3.86 & 3.61 \\ 
\rowcolor{verylightgray}
9596 & 5026 & -19.78 & 9309.24 & 596.16 & 138.88 & 8.99 & 3.03 \\ 
11404 & 5746 & -20.1 & 100.9 & 1932.33 & 1.14 & 21.74 & 3.6 \\ 
\rowcolor{verylightgray}
12704 & 6280 & -20.03 & 1941.63 & 292.62 & 22.22 & 3.81 & 3.31 \\ 
11074 & 5616 & -20.03 & 560.66 & 213.62 & 7.34 & 2.8 & 3.7 \\ 
\rowcolor{verylightgray}
9766 & 5096 & -19.43 & 971.32 & 219.52 & 13.33 & 3.74 & 3.29 \\ 
11592 & 5833 & -19.85 & 685.39 & 163.08 & 14.68 & 3.5 & 3.59 \\ 
\rowcolor{verylightgray}
12439 & 6184 & -19.88 & 5841.01 & 444.68 & 87.62 & 6.72 & 3.69 \\ 
11127 & 5628 & -18.88 & 2367.73 & 315.82 & 62.12 & 12.78 & 3.19 \\ 
\rowcolor{verylightgray}
12575 & 6234 & -19.01 & 25349.49 & 833.42 & 1117.86 & 119.07 & 3.69 \\ 
11328 & 5665 & -20.47 & 914.45 & 1559.4 & 30.53 & 52.07 & 3.82 \\ 
\rowcolor{verylightgray}
11149 & 5645 & -20.03 & 3232.13 & 391.39 & 43.51 & 5.6 & 3.11 \\ 
14891 & 7207 & -21.53 & 701.74 & 146.73 & 14.99 & 3.42 & 5.5 \\ 
\rowcolor{verylightgray}
10699 & 5447 & -19.71 & 778.12 & 290.28 & 7.29 & 3.86 & 2.99 \\ 
13365 & 6526 & -19.51 & 1424.6 & 301.88 & 25.31 & 5.44 & 3.06 \\ 
\rowcolor{verylightgray}
14541 & 29103 & -19.23 & 5399.7 & 417.07 & 118.07 & 9.6 & 3.7 \\ 
13084 & 6433 & -19.92 & 66.4 & 180.43 & 0.58 & 1.58 & 2.92 \\ 
\hline
 \end{tabular}
\end{table*}

\begin{table*}
\contcaption{}
\setlength\extrarowheight{3pt}
\centering
\begin{tabular}{| c | c | c | c | c | c | c | c | c |}
\hline

ID & ID & $M_{UV}$ & Flux [3-Kron] & Flux $\sigma$ & rest-EW & rest-EW $\sigma$ & z \\ 
(Skelton) & (Guo) & (F814W) &  ($\times 10^{-20}$ erg cm$^{-2}$ s$^{-1}$ \AA$^{-1}$)  &  ($\times 10^{-20}$ erg cm$^{-2}$ s$^{-1}$ \AA$^{-1}$)  & (\AA) & (\AA) &   \\ 
\hline
\hline
\rowcolor{verylightgray}
13532 & 6616 & -19.15 & 1856.91 & 284.91 & 42.72 & 7.73 & 3.02 \\ 
15546 & 7538 & -20.52 & 2299.63 & 348.04 & 51.07 & 9.25 & 5.52 \\ 
\rowcolor{verylightgray}
16198 & 7847 & -20.2 & 2190.42 & 317.52 & 30.07 & 4.37 & 3.14 \\ 
14809 & 7154 & -20.46 & 1705.43 & 218.19 & 15.62 & 2.18 & 3.65 \\ 
\rowcolor{verylightgray}
15130 & 7304 & -20.97 & 96.06 & 128.01 & 1.17 & 1.56 & 5.1 \\ 
14703 & 7151 & -19.93 & 1089.17 & 197.41 & 21.13 & 6.35 & 4.09 \\ 
\rowcolor{verylightgray}
17385 & 8493 & -20.13 & 440.18 & 180.15 & 5.57 & 3.03 & 3.07 \\ 
17741 & 8702 & -20.07 & 1556.26 & 247.15 & 7.65 & 2.88 & 3.31 \\ 
\rowcolor{verylightgray}
16269 & 7896 & -19.17 & 3341.44 & 300.64 & 83.96 & 7.9 & 3.32 \\ 
18872 & 9340 & -19.48 & 4635.43 & 412.64 & 78.83 & 7.05 & 3.0 \\ 
\rowcolor{verylightgray}
19538 & 9735 & -19.51 & 836.4 & 151.79 & 22.1 & 4.7 & 4.26 \\ 
17356 & 8485 & -19.54 & 5972.33 & 488.59 & 67.19 & 5.63 & 3.0 \\ 
\rowcolor{verylightgray}
17612 & 8621 & -19.44 & 3032.74 & 320.9 & 42.72 & 5.48 & 3.29 \\ 
16398 & 7947 & -20.33 & 1108.53 & 201.67 & 10.54 & 1.93 & 3.49 \\ 
\rowcolor{verylightgray}
18576 & 29778 & -19.72 & 454.82 & 178.97 & 22.37 & 8.86 & 4.43 \\ 
19390 & 9653 & -20.12 & 2314.72 & 266.92 & 31.71 & 3.84 & 3.57 \\ 
\rowcolor{verylightgray}
19717 & 9858 & -19.64 & 7681.29 & 451.84 & 193.86 & 13.46 & 3.42 \\ 
18773 & 9266 & -19.41 & 1715.65 & 255.01 & 52.61 & 8.06 & 3.55 \\ 
\hline

 \end{tabular}
\end{table*}



\bsp	
\label{lastpage}
\end{document}